\documentclass{aastex}
\usepackage{spr-astr-addons}

\RequirePackage{color}

\newcommand{\emaila}{husm@sdu.edu.cn}

\begin{document}

\title{variability and spectral variation of 3C~66A}
\shorttitle{variability of 3C~66A}
\shortauthors{Hu et al.}

\author{Shao Ming Hu\altaffilmark{1,2}}
 \and
\author{Jianghua Wu\altaffilmark{3}}
 \and
\author{H. Y. Guo\altaffilmark{4}}
 \and
\author{X. Zhou\altaffilmark{3}}
\and
\author{X. Zhang\altaffilmark{5}}
\and
\author{Y. G. Zheng\altaffilmark{5}}

\altaffiltext{1}{Shandong Provincial Key Laboratory of Optical
Astronomy and Solar-Terrestrial Environment, Shandong University at
Weihai, 180 Cultral West Road, Weihai, Shandong, 264209, China
 Email: \emaila} \altaffiltext{2}{Max-Planck Institute for Extraterrestrial
physics, Munich, 85741, Germany} \altaffiltext{3}{National
Astronomical Observatories of Chinese Academy of Sciences, Beijing,
10012, China} \altaffiltext{4}{School of Computer Science, Harbin
Institute of Technology at Weihai, Weihai, China}

\altaffiltext{5}{Department of Physics, Yunnan Normal University,
Kunming, Yunnan, China}

\begin{abstract}
3C~66A was monitored by the BATC
(Beijing-Arizona-Taipei-Connecticut) telescope from 2005 to 2008,
1994 observations were obtained on 89 nights. Detailed research and
analysis was performed on these observations in this paper. A long
term burst occurred in the whole light curve. No intra-day
variability was claimed in our campaign by intra-night light curve
analysis. Time lag of shorter wavelenth preceding longer wavelength
was shown by correlation analysis. The results showed that the
optical spectral shape turned flatter when the source brightened,
and the spectral variability indicator was bigger on shorter
time-scale as determined by the color indices variation analysis.
\end{abstract}



\section{Introduction}

3C~66A is a classic blazar object, and blazars are the most extreme
subclass of the active galactic nuclei. They show highly variable
non-thermal continuum emission ranging from radio up to X-ray and
often to $\gamma$-ray frequencies on different time scales, and they
exhibit strong polarization from radio to optical wavelength
\citep[e.g.,][]{maraschi,urry,hartman,gupta,raiteri2008,villata2009,bauer}.
Blazars can be classified into BL Lac objects (BL~Lacs) and
flat-spectrum radio quasars (FSRQs), based on their emission line
features \citep{landt}. The former one has a featureless optical
continuum or has very weak emission or absorption lines, while the
latter has some strong and broad emission lines. Studies of
broadband variability, spectral variability and time lags between
flux variations at different wavebands play an important role to
understand the radiation and variation mechanism and physical
structure of AGNs \citep[See][and reference therein]{wagner,ulrich}.

3C~66A is classified as a low-frequency peaked BL Lac objects (LBL),
according to the peak frequency of synchrotron emission
\citep{padovani,giommi1999}. Long-term optical variability or
optical spectral variability have been studied by lots of
investigators \citep{fan,damicis,vagnetti,hus}: The study of
\citet{fan} showed that the variability amplitude increased with
frequency, and all these investigations showed the optical spectra
turns bluer/flatter when the source brightens, although this
tendency was strongly dependent on three data points in the study by
\citet{fan}, but this tendency was confirmed by many other authors
\citep[e.g.,][]{gu,dai,rani}. Short term variability and color
variations have been investigated by many authors
\citep[e.g.,][]{takalo,gu,rani}: All their results showed short term
variability, and the OJ-94 project on 3C~66A \citep{takalo} showed
many intensive light curves; Significant positive correlation
between V-I index and R magnitudes was detected by \citet{gu};
Significant positive correlation between V-R, R-I indices and V
magnitudes were shown by \citet{rani} but not for the other two
color indices. Intra-day variability has also been previously
reported by \citet{carini},\citet{takalo} and \citet{diego}, but
some observations did not reveal intra-day variability
\citep[see][]{miller,takalo1992,xie}. 3C~66A has been intensively
monitored by two WEBT (Whole Earth Blazar Telescope) compaigns
\citep[see][]{bottcher05,bottcher09}. The first WEBT campaign was
from 2003 July through 2004 April. Their results revealed intra-day
variability within a 2-hour time-scale. During this campaign their
study indicate a positive hardness-intensity correlation for low
optical fluxes, which did not persist at higher state. B-R color
index seemed to reach its minimum values several days prior to the B
and R flux peaks. The second WEBT campaign covered the autumn and
winter of 2007--2008. Evidence of rapid intra-day variability was
revealed. There was no systematic spectral variability in the high
state (R$\leq $14.0) detected in this campaign, and there was no
evident time lags among different optical bands by their analysis.

In order to investigate the variability, spectral variability and
the time lag properties of this object, we monitored 3C~66A with
BATC $60/90cm$ Schmidt telescope from 2005 January 26 to 2008
October 16. The observations and data reduction are described in
section 2, the variability, optical spectral variability properties
and time lags are presented in section 3, and section 4 gives a
summary and discussion.

\section{Observations and data reduction}

Observations were carried out with a 60/90cm f/3 Schmidt telescope,
which is located at Xinglong Station of National Astronomical
Observatories of China. Images were recorded by a Ford Aerospace
$2\,048\times2\,048$ CCD camera, mounted at the focus plane before
the early of 2006. The pixel size of this CCD is 15 microns and the
field of view is $58'\times58'$, resulting in a resolution of
1.7\arcsec/pixel. A new E2V $4\,096\times4\,096$ CCD was in
commission from the early of 2006. The pixel size of the new CCD is
12 microns, so it results in a spatial resolution of
1.3\arcsec/pixel. The telescope is equipped with a 15-colour
intermediate-band photometric system, covering a wavelength range of
3\,000--10\,000 \AA. The telescope and the photometric system are
mainly used to carry out the BATC survey \citep{zhou2005}. From 2005
January 26 to 2006 November 23, observations of 3C~66A were
performed with the BATC e, i and m bands in cyclic mode, whose
central wavelengths are 4485\AA, 6685{\AA} and 8013\AA,
respectively. After that it was observed with BATC c, i and o
filters in cyclic mode, so most of our observations are in c, i, and
o bands. The central wavelengths of c and o band are 4206{\AA} and
9173\AA, respectively. The field of view of the central
$512\times512$ pixels is large enough to cover the object and its
comparison stars, so only the central $512\times512$ pixels are read
out for blazar monitoring to shorten the readout time. For a good
compromise between photometric precision and temporal resolution, we
took 30--600 seconds as the exposure time for different filters,
according to the weather condition and moon phase.

All images were processed with a pipeline data reduction procedure
including bias subtraction, flat fielding, position calibration,
aperture photometry and flux calibration. The adopted aperture was
8\arcsec. The inner and outer radii of sky annulus were set to
9.1\arcsec and 13\arcsec, respectively. Comparison stars A, B, C1
and C2 \citep[from][]{gonzalez} were used for differential
photometry during the data reduction. Their BATC magnitudes were
obtained by observing them and three BATC standard stars, HD~19445,
HD~84937 and BD+17d4708 on photometric nights. The BATC magnitudes
of the standard stars were described by \citet{yan}. The extinction
coefficients and zero points were obtained by the observations of
these three standard stars, then the instrument magnitudes of the
comparison stars were transferred to BATC magnitudes. These
magnitudes and their errors were listed in Table~\ref{standard}.
During the calibration, the instrumental magnitudes of 3C~66A and
all the comparison stars were extracted first, then the object
brightness was calculated to be the average of differential
photometry magnitudes of the object relative to comparisons A, B and
C1. C2 was chosen as the check star, which was used to verify the
accuracy of our measurements. For each image, the difference between
the instrumental magnitude of C2 and the average instrumental
magnitude of A, B, and C1 was obtained, then the difference between
this value and its average over the whole night was defined as
$S_{x}$ ($x$ indicates the BATC band c, e, i, m and o.) for
quantifying the accuracy of the measurements. The details on data
reduction were described by \citet{yan}, \citet{zhou2003} and
references therein. The BATC magnitudes can also be transferred to
standard Johnson-Cousins magnitudes by the relations obtained by
\citet{zhou2003}. Covering four years from 2005 January 26 to 2008
October 16, 3C~66A was observed on 89 nights, during which
observation time ranged from half an hour to 6 hours from night to
night, because of weather condition or other observation projects.
In order to keep the accuracy of the following study, all the
observations with photometry error bigger than 0.05 mag were
discarded, so the error bars of the magnitude were not plotted in
the following figures for clarity. Finally, 1994 observations from
BATC were produced. All the observations are listed in
Table~\ref{observations}, which gives an example of the photometry
results in print version. Complete content is only available in
electronic version. The universal date of observations (Col.~1) is
followed by the universal time at the middle of exposure (Col.~2),
Julian date at the middle of exposure (Col.~3), BATC band (Col.4),
exposure time (Col.~5), BATC magnitude corresponding to the BATC
band (Col.~6), photometry error $E_{x}$ (Col.~7) and $S_{x}$, which
indicates the confidence of the observation (Col.~8).

\begin{table*}
\small
\caption{BATC magnitudes and errors of the standard stars.}
\label{standard}
\begin{tabular}{ccccc}
\hline
Band &  A  &     B  &   C1   & C2 \\
\hline
 c & 14.060$\pm0.013$ & 16.020$\pm0.037$ & 12.865$\pm0.011$ & 15.497$\pm0.025$ \\
 e & 13.810$\pm0.010$ & 15.157$\pm0.018$ & 12.854$\pm0.009$ & 14.581$\pm0.013$ \\
 i & 13.516$\pm0.009$ & 14.389$\pm0.013$ & 12.932$\pm0.008$ & 13.741$\pm0.010$ \\
 m & 13.463$\pm0.012$ & 14.164$\pm0.017$ & 12.998$\pm0.010$ & 13.484$\pm0.013$ \\
 o & 13.468$\pm0.014$ & 14.027$\pm0.020$ & 13.003$\pm0.012$ & 13.382$\pm0.015$ \\
\hline
\end{tabular}
\end{table*}

\begin{table*}
\caption{Photometry results.} \label{observations}
\begin{tabular}{cccccccc}
\hline
Date(UT) &  Time(UT)  &  JD     & Band &  Exp   &  Mag  & $E_{x}$  & $S_{x}$ \\
yyyymmdd & hh:mm:ss.s & (day)   &      &   (s)  & (mag) & (mag)    &(mag) \\
\hline
20061126 & 14:01:56.0 & 2454066.08468 & c & 240 & 15.292 & 0.022 & -0.024 \\
20050129 & 10:43:15.0 & 2453399.94670 & i & 180 & 14.742 & 0.018 &  0.002 \\
20061126 & 14:26:02.0 & 2454066.10141 & o & 240 & 14.128 & 0.033 & -0.002 \\
20050129 & 10:38:01.0 & 2453399.94307 & e & 300 & 15.241 & 0.036 & -0.002 \\
20051210 & 15:19:35.0 & 2453715.13860 & m & 450 & 14.437 & 0.018 & -0.010 \\
\hline
\end{tabular}
\end{table*}

\section{Variability and color variability}

Variability and intra-day variability were analysed in detail using
these observations. Color indices variability and time lags were
studied adding the data taken and assembled by WEBT
\citep{bottcher05}. Part of our observations overlapped with the
second WEBT campaign on this source; the light curves during the
overlapped period have been shown in Fig.~2 in \citet{bottcher09},
and the variation agreed with the observations of WEBT campaign. But
no comprehensive analysis was done on these data, because the BATC
photometry filters are different from the filter system used by
other participators.

\subsection{Variability}

\begin{figure}
\centering
\includegraphics[width=\columnwidth]{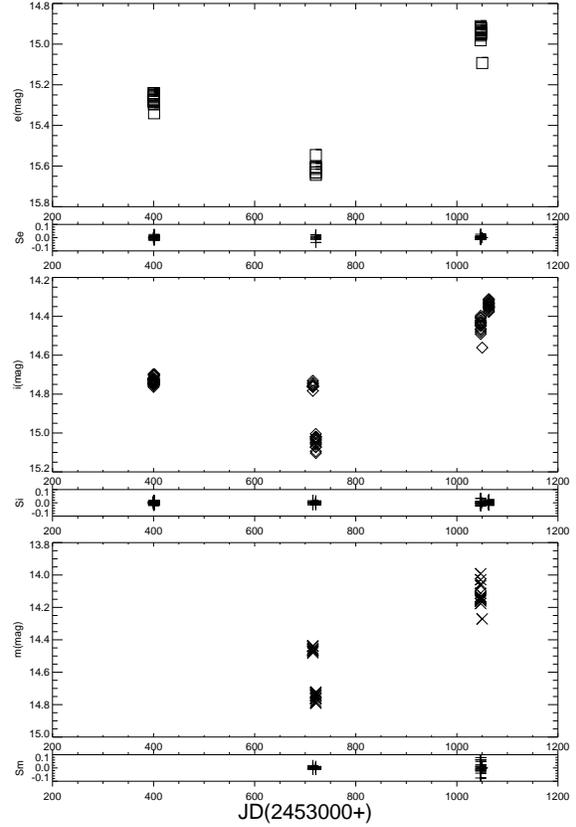} \caption{Light curves of 3C~ 66A in
BATC e, i and m bands from 2005 January 26 to 2006 November 23. The
squares, diamonds and crosses show the light curve in e, i and m
band, respectively. Plus signs in the narrow panels plot the
deviation of $S_{x}$.} \label{longlc2}
\end{figure}

\begin{figure}
\centering
\includegraphics[width=\columnwidth]{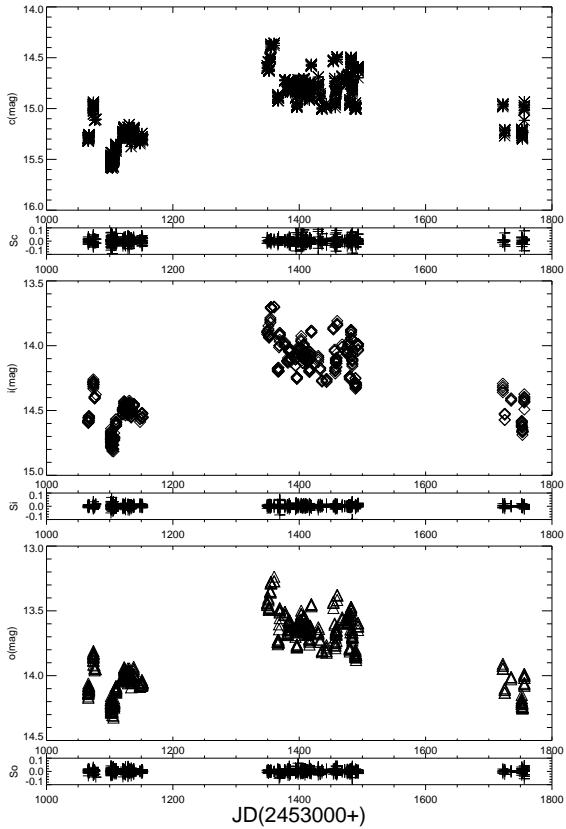} \caption{Light curves of 3C~ 66A in
BATC c, i and o band from 2006 November 26 to 2008 October 16. The
asterisks, diamonds and triangles demonstrate the light curve in c,
i and o band, respectively. Plus signs in the narrow panels show the
deviation of $S_{x}$.} \label{longlc1}
\end{figure}

Fig.~\ref{longlc2} shows the light curves of 3C~66A in the BATC e, i
and m bands from 2005 January 26 to 2006 November 23, and
Fig.~\ref{longlc1} shows the light curves of 3C~66A in c, i and o
bands from 2006 November 26 to 2008 October 16. The narrower panels
below the light curves present the variation of $S_{x}$, which
indicates the confidence of the observations. One can see that all
of them are less than 0.1 mag, and most of them are within 0.05 mag,
which is almost at the same level as the photometry errors. Large
amplitudes of variation were detected in our monitoring; the
variation amplitude is 0.731, 0.792, 0.799 mag in e, i, and m band
in Fig.~\ref{longlc2}, respectively. A big burst occurred during the
second monitoring phase, and lots of fast oscillations are
superimposed on the long term burst. The variation magnitude is
1.231, 1.406, 1.087 mag in c, i and o band, respectively, but it is
a pity we could not monitor it during the two gaps because of the
observing season or other observation projects.

The monitoring lasted several hours on many nights, which provided
us a good opportunity to study the intra-day variability for 3C~66A.
We took the criterion presented by \citet{jang} to quantitatively
claim whether the object was variable or not within one night. They
defined a parameter C as $C=\sigma_{T}/\sigma$, where $\sigma_{T}$
is the standard deviation of the light curve, and $\sigma$ is the
standard deviation of $S_{x}$. Because $S_{x}$ presents the
variation of the comparison stars, $\sigma$ can describe the
confidence of the observations better. $C$ is taken as the
confidence parameter of variability. The source will be claimed to
be variable at 99\% confidence level when $C\geq2.576$. We will say
the object is variable only when $C\geq2.576$ at least in two bands
during one night if it was monitored in three or more bands
\citep{jang,stalin}. All the intra-night light curves continuously
observed longer than 2 hours in the same band within the same night
were analyzed, and all the results were listed in Table~\ref{inov}.
The observation universal date (Col.~1) is followed by the BATC band
(Col.~2), number of observations (Col.~3), the duration of the light
curve in unit of hour (Col.~4), the value of C as defined above
(Col.~5), and a label which indicates the source variable (1) or not
(0) according to the criteria described above (Col.~6). The
variability amplitude reached 0.1mag and even to 0.2mag on a few
nights, but according to the criteria by \citet{jang}, no intra-day
variability was claimed on these 15 nights in our campaign. The same
result has been reported by \citet{miller}, \citet{takalo1992} and
\citet{xie} in their observations, but \citet{diego} have detected
rapid intra-day variability in the infrared band.

\begin{table}
\centering \caption{Intra-day variability analysis results.}
\label{inov}
\begin{tabular}{ccccrrcrcr}
\hline
  Date   & Band&      N    &   H(h)   &   C    & V/N\\
\hline

20061107 &  e  &     17    &   3.5    &  1.560 &  0\\
20061107 &  i  &     17    &   3.5    &  1.644 &  0\\
20061107 &  m  &     17    &   3.6    &  1.063 &  0\\
20061126 &  c  &     12    &   2.7    &  0.819 &  0\\
20061126 &  i  &     14    &   3.1    &  1.411 &  0\\
20061126 &  o  &     14    &   3.1    &  2.399 &  0\\
20061127 &  c  &     12    &   2.4    &  1.431 &  0\\
20061127 &  i  &     12    &   2.4    &  1.397 &  0\\
20061127 &  o  &     12    &   2.4    &  1.642 &  0\\
20061204 &  c  &     14    &   3.3    &  0.958 &  0\\
20061204 &  i  &     15    &   3.3    &  1.388 &  0\\
20061204 &  o  &     17    &   3.7    &  1.581 &  0\\
20070101 &  c  &     24    &   5.6    &  1.122 &  0\\
20070101 &  i  &     24    &   5.6    &  2.078 &  0\\
20070101 &  o  &     24    &   5.6    &  1.887 &  0\\
20070102 &  c  &     22    &   5.7    &  1.283 &  0\\
20070102 &  i  &     22    &   5.7    &  1.408 &  0\\
20070102 &  o  &     22    &   5.7    &  1.870 &  0\\
20070103 &  c  &     18    &   5.5    &  0.621 &  0\\
20070103 &  i  &     18    &   5.5    &  1.682 &  0\\
20070103 &  o  &     18    &   5.5    &  1.427 &  0\\
20070106 &  c  &     12    &   4.9    &  0.943 &  0\\
20070106 &  i  &     13    &   5.2    &  1.281 &  0\\
20070106 &  o  &     13    &   5.2    &  1.669 &  0\\
20070109 &  c  &     17    &   5.0    &  1.161 &  0\\
20070109 &  i  &     17    &   5.0    &  1.148 &  0\\
20070109 &  o  &     17    &   5.0    &  1.395 &  0\\
20070120 &  c  &     17    &   3.9    &  1.139 &  0\\
20070120 &  i  &     17    &   3.9    &  1.357 &  0\\
20070120 &  o  &     17    &   3.9    &  0.894 &  0\\
20070122 &  c  &     13    &   2.8    &  0.491 &  0\\
20070122 &  i  &     13    &   3.1    &  2.111 &  0\\
20070122 &  o  &     13    &   3.1    &  1.946 &  0\\
20070124 &  c  &     15    &   3.3    &  1.103 &  0\\
20070124 &  i  &     15    &   3.3    &  1.852 &  0\\
20070124 &  o  &     15    &   3.3    &  1.119 &  0\\
20071010 &  c  &     13    &   2.1    &  0.576 &  0\\
20071010 &  i  &     13    &   2.1    &  0.808 &  0\\
20071010 &  o  &     13    &   2.1    &  0.774 &  0\\
20080118 &  c  &     13    &   3.4    &  0.860 &  0\\
20080118 &  i  &     13    &   3.4    &  0.722 &  0\\
20080118 &  o  &     13    &   3.4    &  0.668 &  0\\
20080123 &  c  &     14    &   3.8    &  1.048 &  0\\
20080123 &  i  &     14    &   3.7    &  2.328 &  0\\
20080123 &  o  &     14    &   3.7    &  2.241 &  0\\

\hline
\end{tabular}
\end{table}

\subsection{Spectral variability}

The correlation between the brightness and the spectral shape (color
index can be used to trace it in the optical band) of blazars has
been investigated by many authors. Some research showed that the
spectral shape became flat when the brightness increased, such as
results of BL Lac from \citet{racine} and \citet{vagnetti}, result
of OJ~287 from \citet{gear}. But the opposite correlation has also
been reported \citep[e.g.,][for PKS~0736+017]{ramirez}.
\citet{gu},\citet{hus}, and \citet{rani} showed that this relation
may be different for different subclasses of blazar.
\citet{bottcher05} concluded the correlation was different when the
source was in different states. \citet{villata2004a} and \citet{hu}
claimed the correlation level was different on different time-scale.
The observation strategy of this 4-year monitoring can produce dense
observations in different bands, which makes it possible to
calculate the quasi-simultaneous color index to analyse the color
variability and the correlation between brightness and color
indices.

\begin{figure}
\centering
\includegraphics[width=\columnwidth]{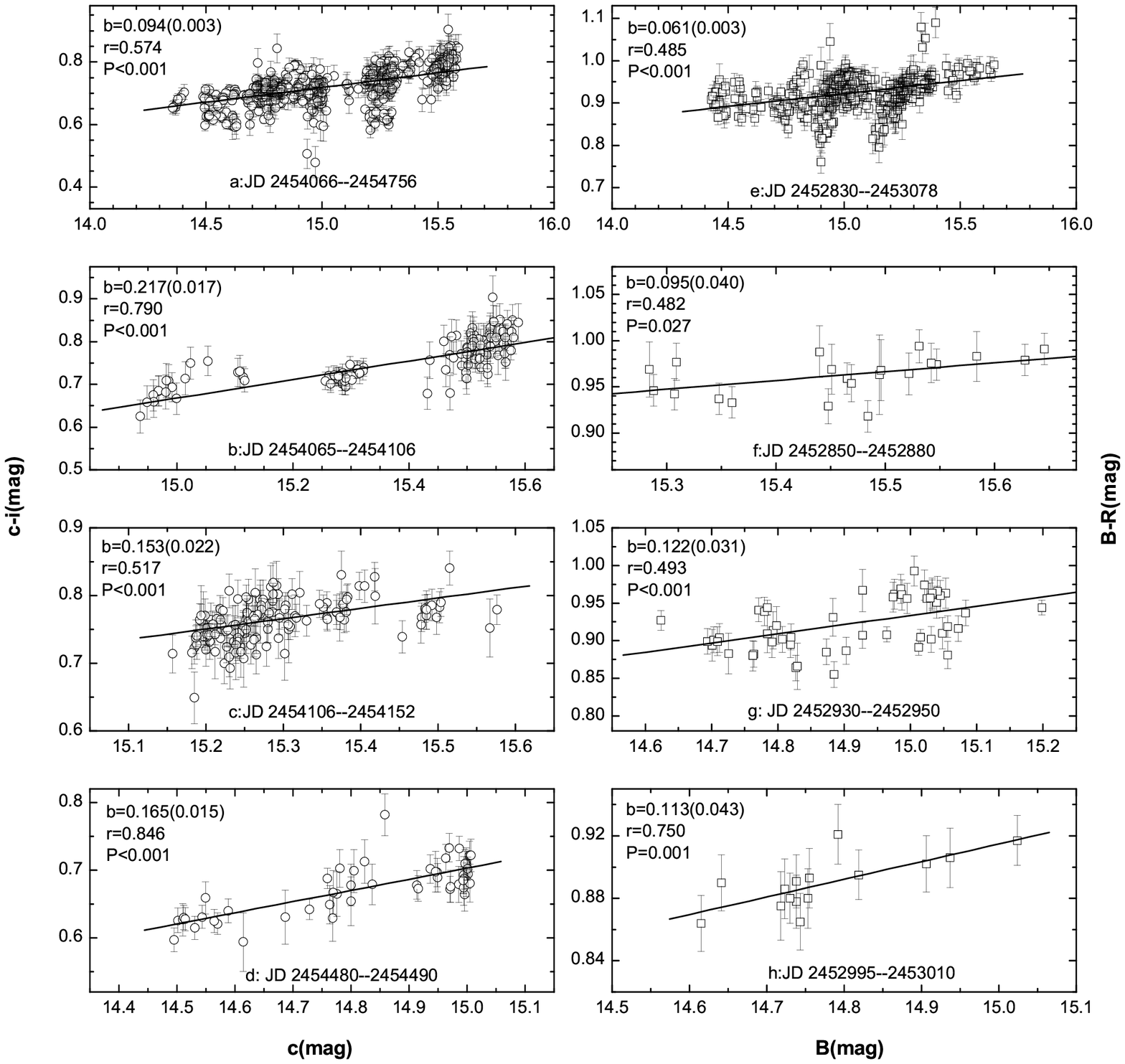} \caption{The correlation between the
color index and the brightness. Circles in the left four panels
display the relationship between c and c-i, and squares in the right
four panels show the correlation between B and B-R. The solid lines
are the best linear fit. The errors of c and B are less than 0.05
and 0.08 mag, respectively, and they are not plotted for clarity.}
\label{colormag}
\end{figure}

In order to study the color variability of 3C~66A, observations in
the c and i bands were chosen. WEBT has made huge contributions to
the study of blazars \citep[e.g.,][and references
therein]{raiteri2007,raiteri2008,villata2009}. The optical
observations in B and R bands of the extensive monitoring of the
first WEBT campaign \citep{bottcher05} were also used for the
analysis in this paper. We chose observations in these bands because
they were the most densely sampled bands in these two campaigns.
Color indices were calculated using different band magnitudes taken
within 30 minutes, and most of them are within 15 minutes. The error
of the color index was calculated according to the rules of error
propagation. The relation between the brightness and the color
indices is demonstrated in Fig.~\ref{colormag}. The correlation
between c and c-i color index in the BATC campaign is given in the
left four panels, and the right four panels show the correlation
between B and B-R color index in the WEBT campaign. The top two
panels illustrate the long term correlation covering each whole
campaign, but the bottom six panels display the short flares
identified from the long term light curves. The solid lines are the
best Peterson linear fit. The fitted line slope b, which is taken as
the color variability indicator, the error of b, the linear
correlation coefficient r and the probability that r is zero are
given in each panel. The observation time interval of each panel is
labeled at the bottom of each panel. A positive correlation between
brightness and the color index is found for all cases. The value of
b for the BATC long term light curve is 0.094, and those for the
three short flares are 0.217, 0.153, 0.165. The value of b for the
WEBT long term light curve and the three short flares are 0.061,
0.095, 0.122 and 0.113, respectively. The probability that r is zero
for each linear fit is less than 0.03. We can see that the color
variability indicators of the short flares are all larger than that
of long term light curve. The result is the same if we discard the
observations in the high state ($R\leq 14.0$) \citep[see Sec.~3
of][]{bottcher05} in the WEBT compaign, because the difference of b
value is very small. So we conclude that the color variability
indicator is larger on shorter time-scale for 3C~66A. This kind of
tendency has been reported by \citet{villata2004a} and \citet{hu}
for BL Lac.

\subsection{Time lag}

\begin{figure*}
\centering
\includegraphics[width=\textwidth]{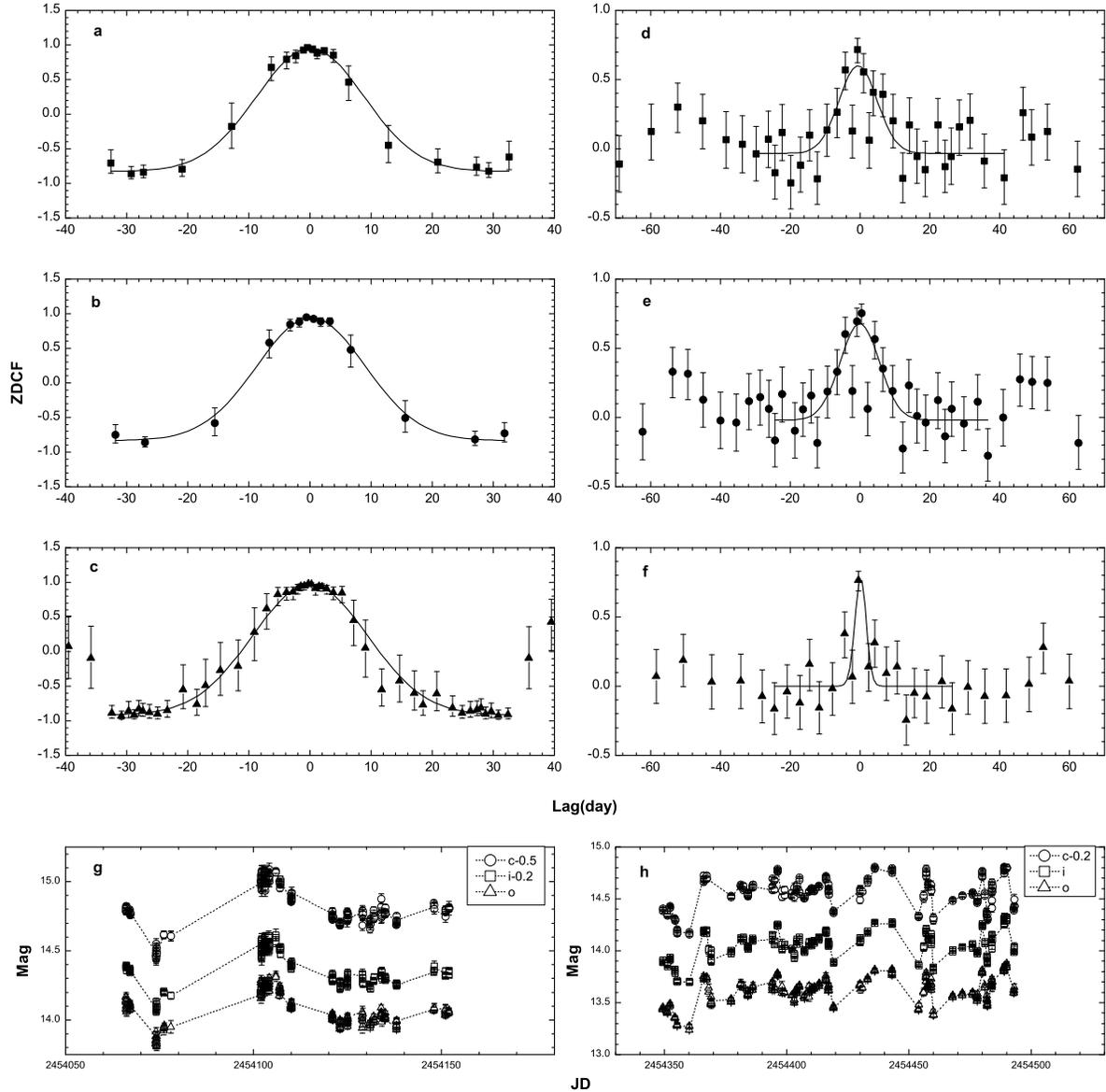} \caption{Time lags between flux variations at different
optical bands by ZDCF. Positive lag means long wavelength emission
precedes that of short wavelength. Filled squares display the time
lag between the light curve of o and c, filled circles show the ZDCF
between i and c, and filled triangles illustrate the ZDCF between o
and i. Panels a, b, c, g illustrate the result of jd1, and panels d,
e, f, h display the results of jd2. Solid lines are the best gauss
fit for the central part of data. The bottom two panels display the
light curves.} \label{olag}
\end{figure*}

It can be seen that the shape of the light curves of c, i and o are
in agreement with each other, so the z-transform discrete
correlation function \citep[ZDCF,][]{alexander} was applied to the
light curves. The ZDCF was used here as an estimation of the cross
correlation function to calculate the time lags between different
light curves. It applies z transformation to the correlation
coefficients and uses equal population bins rather than the equal
time bins in discrete correlation function \citep[DCF,][]{edelson}.
One advantage of this technique is that it allows direct estimation
of the uncertainties on the lag without more assumption-dependent
Monte Carlo simulations. It has been shown in practice that the
calculation of the ZDCF is more robust than that of the DCF when
applied to sparsely and unequally sampled light curves
\citep[e.g.,][]{edelson96,roy}. The whole BATC campaign data can be
used to detect the time lag, but the bump of the ZDCF curve is very
wide and it is hard to claim a short time lag, because there are
several big gaps in the monitoring. We analyzed the time lag between
different optical bands with more continuous observations during two
periods, one is from JD 2454066 to 2454153 (jd1), another is from JD
2454349 to 2454494 (jd2). The results are shown in Fig.~\ref{olag}.
The top two panels display the ZDCF between the light curve of o and
c, the second two panels from the top show the ZDCF between o and i,
the third two panels from the top illustrate the ZDCF between i and
c and the bottom two panels show the translated light curves of c,
i, o band, whose brightness variation corresponds with each other.
The left panels and right panels display the results of jd1 and jd2,
respectively. All the results are listed in Table~\ref{tbolag},
$ZDCF_{p}$ is the peak value of ZDCF, $\tau_{p}$ is the time lag
corresponding to $ZDCF_{p}$, $\tau_{G}$ is the time lag
corresponding to the center of the best gauss fit of the central
part of ZDCFs and time lags, and Lag (last column) is the average of
$\tau_{p}$ and $\tau_{G}$. Time lags from jd1 are in the same level
with those by \citet{bottcher09}, But we can not get a consistent
result from jd2.

\begin{figure}
\centering
\includegraphics[width=\columnwidth]{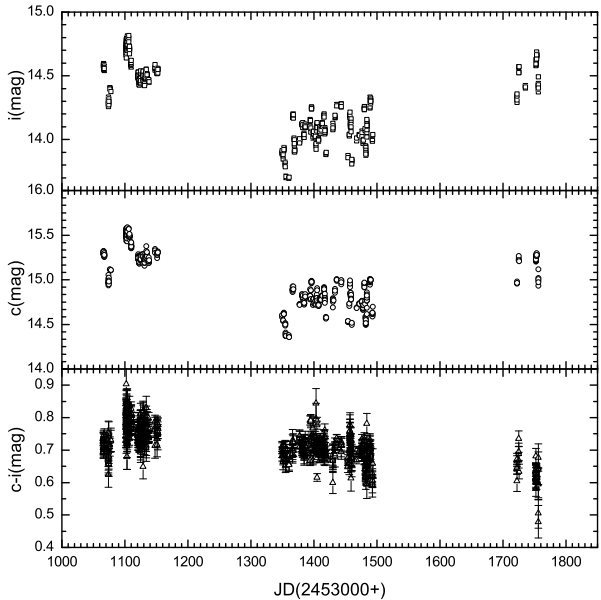} \caption{The light curves and the
color index light curve. Squares, circles and triangles illustrate
the variation of i, c and c-i, respectively.} \label{colorlc}
\end{figure}

\begin{figure*}
\centering
\includegraphics[width=\textwidth]{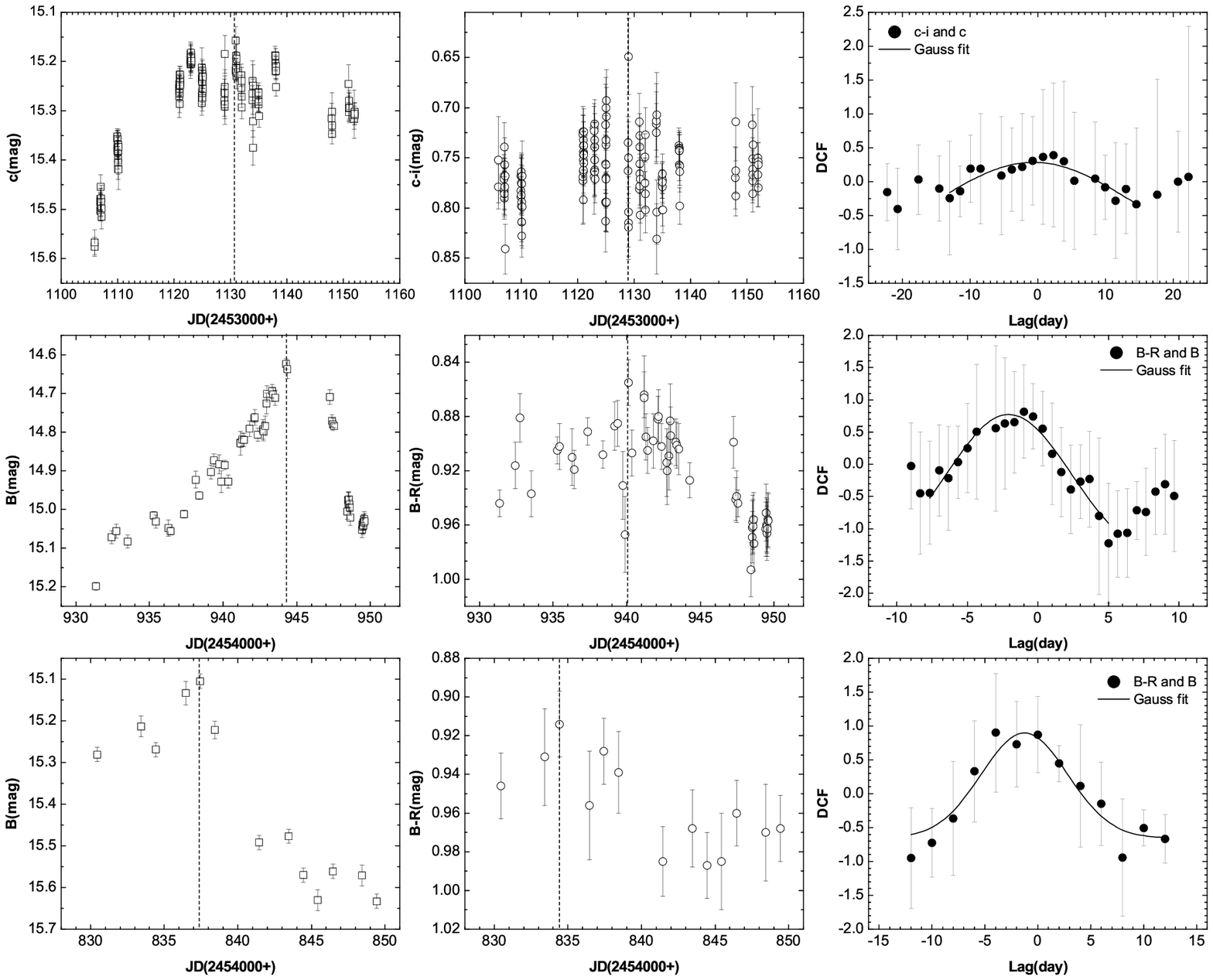} \caption{The time lag between the
color index and the brightness. Negative time lag means color index
variation precedes the variation of light curve. Squares and circles
display the light curve of c(B) and the color variability of
c-i(B-R), respectively. Filled circles plot the time lag between the
color index c-i(B-R) and the magnitude c(B). The solid lines are the
best gauss fit, and the dash lines label the peak of the curves.}
\label{colorlag}
\end{figure*}

Fig.~\ref{colorlc} shows the c and i light curves and the color
index variation curve. While the variation of c is consistent with
the variation of i, the long term color index variation does not
show a strong correlation with the light curves by the discrete
correlation function \citep{edelson} analysis. But the minimum of
the color index precedes the peak of the light curve by 2.07 days
during the short flare from JD 2454106 to 2454152. The result is
illustrated on the top panels of Fig.~\ref{colorlag}. This
precedence could not be seen obviously in Fig.~\ref{colorlc} because
of the sparse and discontinuous observations during our campaign, so
continuous observations by many telescopes located at positions with
different longitude and latitude are crucial to study the spectral
variation of blazars. The DCF analysis result was noisy and did not
show a strong correlation between c-i and c during this flare; but
the Gauss fit was applied to the bump of DCF, and the fitted center
is $-0.63\pm3.84$ day, which implies that the color index precedes
the light curve of c band. This result agree with the result by
\citet{bottcher05}, and two short flares from \citet{bottcher05} are
illustrated in Fig.~\ref{colorlag}. The middle panels show the
result of the flare from JD 2454930 to 2454950 and the bottom panels
display the flare from JD 2454820 to 2454850. The B-R color index
precedes the light curve of B band by 4.16 and 2.99 days,
respectively. DCF analysis was also applied to these two flares;
both of the peaks of DCF are higher than that of BATC because of
denser and more continuous sampling. The lags are $-1.0$ and $-4.0$
days corresponding to the peak of the DCF, and the gauss fitted
centers are $-2.06\pm0.66$, $-1.2\pm1.16$ days, which indicates that
B-R precedes B. In other words, the optical spectra harden to the
bluest state before it reaches to the brightest state. It maybe
relates to the physical process of outburst.

\begin{table*}
\caption{Time lags between flux variations at different optical
bands by ZDCF.} \label{tbolag}
\begin{tabular}{lcrrrrr}
\hline
Period & Bands &  $ZDCF_{p}$ & $\tau_{p}(day)$  & $\tau_{G}(day)$ & Lag(day) \\
\hline
jd1 & o--c &   0.960 & $ -0.386_{-  0.364}^{+  0.136}$ & $  0.086\pm  0.439$ & $ -0.150$ \\
jd2 & o--c &   0.716 & $ -0.664_{-  0.336}^{+  0.414}$ & $ -0.608\pm  0.986$ & $ -0.636$ \\
jd1 & i--c &   0.979 & $ -0.250_{-  0.210}^{+  0.126}$ & $  0.049\pm  0.259$ & $ -0.101$ \\
jd2 & i--c &   0.765 & $ -0.371_{-  0.629}^{+  0.621}$ & $  0.333\pm  0.602$ & $ -0.019$ \\
jd1 & o--i &   0.947 & $ -0.550_{-  0.450}^{+  0.300}$ & $  0.190\pm  0.494$ & $ -0.180$ \\
jd2 & o--i &   0.753 & $  0.392_{-  0.642}^{+  0.608}$ & $ -0.104\pm  0.868$ & $  0.144$ \\
\hline
\end{tabular}
\end{table*}

\section{Summary and discussion}

3C~66A was monitored by the 60/90cm Schmidt telescope from 2005
January 26 to 2008 October 16 on 89 nights; 1994 observations in 5
BATC bands were obtained. Intra-day variability was analysed by the
criteria presented by \citet{jang} on 15 nights, on which the
duration of observations in all three bands were longer than 2
hours, but no intra-day variability was claimed. One long term burst
occurred during our monitoring; the variability amplitudes in c, i
and o band are 1.231, 1.406, 1.087 mag, respectively. The profile of
the light curves in three bands are consistent with each other, and
time lags between different light curves were analyzed by ZDCF
\citep{alexander}. The results from jd1 are consistent with those
obtained by \citet{bottcher09}. The spectral variability was studied
on long term, and short flare time-scale; the results show that the
spectrum turns bluer/flatter when the source turns brighter. Some
models \citep{chiang,wang} state that spectra of blazars become flat
when they turn bright. The spectral variability indicator is bigger
on shorter time-scale; this phenomenon may provide a constrain on
the spectral variability mechanism. The interesting result that the
minimum of the color index precedes the maximum of brightness by a
few days was also detected in this paper. This means the optical
spectra harden to the bluest state before a major outburst.

In the unified scheme of AGNs, blazars are thought to have the jet
structure well-aligned with our line of sight \citep{antonucci}, and
the jet is believed to originate and be accelerated by the central
supermassive black hole surrounded by an accretion disk. Two broad,
well-defined components of the blazar non-thermal dominated spectrum
can be identified. The low-energy emission is commonly explained by
synchrotron emission from non-thermal electrons in a relativistic
jet \citep{konigl}. Many models have been proposed to explain the
high-energy component; they can be roughly divided into leptonic or
hadronic in origin, depending on whether it is electrons or protons
which are responsible for the high-energy emission. While there are
some hadronic models, which invoke protons as the ultimate source of
high-energy emission \citep[see][and references
therein]{mucke,bottcher07}, the majority of the proposed models are
leptonic models, which assume the high-energy emission comes from
inverse Compton scattering of relativistic electrons on some
low-energy seed photons. The source of the seed photons is still an
open question and many possible origins have been involved, such as
internally produced synchrotron photons \citep[Synchrotron
Self-Compton, see][]{jones,marscher1996}, photons from an external
source, such as accretion disk or diffuse isotropic photons coming
from broad line clouds \citep[External Compton,
see][]{dermer,sikora,celotti}, or combinations thereof
\citep{dermer1997}.

Based on our observational results, we think the variability and the
optical color variation could be reasonably explained by
shock-in-jet models that involve relativistic shocks propagating
outwards \citep[e.g.,][]{marscher1985,wagner}, and which was
discussed by \citet{rani}. The larger flares are expected to be
produced by the emergence and motion of a new shock triggered by
some strong variation in a physical quantity such as velocity,
electron density or magnetic field moving into and through the
relativistic jet. Smaller variations may be nicely explained by
turbulence behind a shock propagating down the jet
\citep{marscher1992}. So jet models can produce different
variability at different colors. In the usual boosted synchrotron
models involving shocks propagating down the jets
\citep[e.g.,][]{marscher1985,marscher2008}, radiation at different
frequencies is produced at different distances behind the shocks.
Typically, higher frequency photons emerge and stop sooner
\citep{valtaoja1992,valtaoja2002}. This maybe introduces the result
that the bluest state precedes the maximum of brightness. And we
expect to see time lag which shows hight frequency flux precedes
that of low frequency. This kind of time lag was detected by us even
not in all cases. During the early phase of a rise in flux, one is
more likely to see a bluer color. However, later observation during
the same flare might show a more enhanced redder band as the bluer
one might have stopped rising or may even have passed its peak.
3C~66A a is typical BL Lac object, whose flux is completely
dominated by jet. The former situation should be more likely to be
seen than the latter, yielding a predominance of bluer-when-brighter
situations which was shown by our observations. The geometry of the
jet may also affect the flux. The direction of the jet varies, which
leads to the variability triggered by the variation of Doppler
factor, as the knots move relativistically on helical trajectories
within a small angle with respect to the observer. This may produce
the long-term mildly chromatic variability \citep{villata2004a,hu},
while the intrinsic shock-in-jet mechanism may give a stronger
chromatic short term variability. This scenario can explain the
result that the spectral variability indicator is bigger on shorter
time-scale. So we suggest the intrinsic shock-in-jet mechanism
adding geometric effects may be the model of the variability of
3C~66A. More dense and long term simultaneous multi-color
observations and studies should be carried out to understand and
constrain the radiation mechanism well.

\section*{Acknowledgments}

We thank the anonymous referee for numerous suggestions that helped
to improve this paper. We owe great thanks to all the BATC staffs
who make great efforts to the observations. We are also grateful to
Ralf Bender, Frank Grupp and Anne Bauer for helpful discussion and
revision. This research is supported by the National Natural Science
Foundation of China and Chinese Academic of Sciences joint fund on
astronomy under project No. 10778619, 10778701 and the foundation of
Shandong university at Weihai (No. 2010ZRYB003, 20080030). This work
is partly supported by Max-Planck Institute for Extraterrestrial
physics, University Observatory of the Ludwig-Maximilians
University. This research has made employment of WEBT data and the
NASA/IPAC Extragalactic Database (NED). Jianghua Wu and X. Zhou are
supported by the Chinese National Natural Science Foundation grants
10633020, 10778714, and 11073032.

\bibliographystyle{spr-mp-nameyear-cnd}

\end{document}